# Local positional and spin symmetry breaking as a source of magnetism and insulation in paramagnetic EuTiO$_3$

Oleksandr I. Malyi[1], Xin-Gang Zhao[1], Annette Bussmann-Holder[2], and Alex Zunger[1,*]

[1]Renewable and Sustainable Energy Institute, University of Colorado, Boulder, Colorado 80309, USA

[2]Max-Planck-Institute for Solid State Research, Heisenbergstr. 1, Stuttgart, 70569, Germany

**Email:** Alex.Zunger@Colorado.edu

**Abstract**

We consider theoretically the paramagnetic phases of EuTiO$_3$ that represent configurations created by two sets of microscopic degrees of freedom (m-DOF): positional symmetry breaking due to octahedral rotations and magnetic symmetry breaking due to spin disorder. The effect of these sets of m-DOFs on the electronic structure and properties of the para phases is assessed by considering sufficiently large (super) cells with the required nominal global average symmetry, allowing, however, the *local* positional and magnetic symmetries to be lowered. We find that tendencies for local symmetry breaking can be monitored by following total energy lowering in mean-field like density functional theory, without recourse for strong correlation effects. While most nominally cubic ABO$_3$ perovskites are known for their symmetry breaking due to the B-atom sublattice, the case of f-electron magnetism in EuTiO$_3$ is associated with A- sublattice symmetry breaking and its coupling to structural distortions. We find that (i) paramagnetic *cubic* EuTiO$_3$ has an intrinsic tendency for both magnetic and positional symmetry breaking, while paramagnetic *tetragonal* EuTiO$_3$ has only magnetic symmetry lowering and no noticeable positional symmetry lowering with respect to low-temperature antiferromagnetic tetragonal phase. (ii) Properly modeled paramagnetic tetragonal and cubic EuTiO$_3$ have a nonzero local magnetic moment on each Eu ion, consistent with the experimental observations of local magnetism in the para phases of EuTiO$_3$ significantly above the Néel temperature. Interestingly, (iii) the local positional distortion modes in the short-range ordered para phases are inherited from the long-range ordered low-temperature antiferromagnetic ground state phase.

# I. Introduction

ABO$_3$ oxide perovskites have attracted significant research interest largely because they represent the rich consequences of many possible configurational arrangements of a few basic microscopic degrees of freedom (m-DOF). The latter include *structural motifs* (rotated, tilted, deformed, or disproportionated BO$_6$ octahedra), *spin motifs* in magnetic configuration, and *dipole* motifs in a ferroelectric configuration. The low-temperature ordered structure of ABO$_3$ can be described theoretically with a sufficient number of m-DOFs needed to capture the *ordered polymorphs,* generally modestly small crystallographic, magnetic, or dipolar unit cells. However, the higher temperature para (elastic, magnetic, electric) phases, lacking long-range order, are inherently more complex and may require non-trivial unit cells that can accommodate the required m-DOFs. Here, we analyze theoretically the electronic and magnetic structure of the para phases of EuTiO$_3$ (ETO) studying positional symmetry breaking due to octahedral rotations and magnetic symmetry breaking due to spin disorder and their effects on the electronic and magnetic properties.

ETO containing an Eu$^{2+}$(4f$^7$) magnetic ion on the A- sublattice of the ABO$_3$ perovskite structure and a nonmagnetic Ti$^{4+}$(d$^0$) ion on the B-atom sublattice offers an interesting case of f-electron antiferro as well as para magnetism. ETO exhibits three principal phases confirmed by crystallographic and magnetic data[1-9] (Table I): below the Néel temperature T$_N$~5.4±0.3 K [6,7,10,11], ETO is in the $\alpha$ phase which is an antiferromagnetic (AFM) insulator, classified crystallographically[6] as tetragonal (I4/mcm) space group. Between T$_N$ and T$_S$=282 K, it is in the $\beta$ phase being a paramagnetic (PM) tetragonal (I4/mcm) insulator[1-6,9], whereas at temperatures T>282 K, the $\gamma$ phase is a paramagnetic cubic (Pm-3m) insulator.[1-6,9] Additional phase transitions were inferred within the temperature range assigned to the tetragonal $\beta$ phase on the basis of the temperature-dependence of birefringence[11-13], X-ray diffraction (XRD) data[9], µSR experiments[9], and dielectric measurements[14]. Most of these observations were made under an applied magnetic field. A more recent XRD analysis of polycrystalline ETO on SrTiO$_3$ at 100 K without an applied magnetic field[9] showed no additional structural transition except those shown in Table I, in agreement with the analysis of pair distribution function[2,8] and other XRD analysis[1-8].

**Table I.** Summary of the three phases α, β, and γ of EuTiO$_3$ in increasing order of temperatures, and their magnetic configuration with corresponding DFT calculations presented in this work. The information on the right-hand side describes the results of the current calculations, including the supercell size used, the existence of positional symmetry breaking in the form of octahedral tilting, and the ensuing formation of a polymorphous network; the type of symmetry distortion modes found in the calculation, the distribution of Eu magnetic moments, and the (generally underestimated) DFT band gap. Other literature DFT results are compared in the main text.

| | Experiment | | | Results of calculations in this work | | | | | |
|---|---|---|---|---|---|---|---|---|---|
| Phase name | Nominal crystal structure | Magnetic config. | Temperature (K) | Supercell size (f.u.) | Octahedral tilting | Structural polym. | Distortion mode | Eu local magnetic moments | DFT band gap (eV) |
| γ | Pm-3m | PM | T > 282 | 32 | yes | yes | $R_5^-$, $T_2$, $DT_5$, $M_2^+$ | 6.88±0.005 | 0.27 |
| β | I4/mcm | PM | ~5.4 < T < 282 | 32 | yes | no | $R_5^-$ | 6.88±0.005 | 0.31 |
| α | I4/mcm | AFM | T < 5.4±0.3 | 4 | yes | no | $R_5^-$ | 6.88 | 0.33 |

*Different literature views on the nature of the microscopic structure of the PM configurations:* PM phases, in general, are defined by the lack of long-range order and having a total zero magnetization. The simplest conceptual realization of these conditions is that *each* site would have a zero moment. This initial view resulted in referring to paramagnetic phases as "*nonmagnetic*" or "*nominally nonmagnetic*".[11,14] Such a nonmagnetic interpretation of paramagnetism predicts, within the density functional theory (DFT) as will be illustrated below, a gapless electronic structure, in contradiction with diffuse reflectance spectra measured at room temperature of EuTiO$_3$, showing an insulating band gap of ~0.8 eV.[15] We note in passing that the scenario of describing a PM as a collection of nonmagnetic sites has been common in the band structure literature for *d-electron perovskites[16-18]* and binary d electron oxides[19]. This view led to the well-known contradiction between the predicted (false) metallic character in PM oxides vs. the observed insulating character, leading to the rise of the electron correlated view of Mott gapping as a solution of this contradiction.[20,21] However, it was recently demonstrated[22-27] that allowing for spatial and spin symmetry breaking leads to proper gapping even in the mean-field band theory.

A more advanced interpretation of the microscopic character of spin disordered PM phases (analogous to chemical disorder in alloys) has allowed non-zero local moments that are disordered in a specific way. For example, in the Disordered Local Moment (DLM) model[28-30] within the single-site (SS) coherent-potential approximation (CPA)[31] it was assumed that the moment and charge on a given site is independent of its local environment, e.g., on how many spin up and how many spin down neighbors coordinate this site. This particular restriction in the CPA DLM leads to a picture that each local moment on a site is identical for all magnetic sites–neglecting charge and moment fluctuations. However, it was later noted that this particular version of a disorder underlying the site-coherent potential approximations leads incorrectly to vanishing electrostatic Madelung energy.[32,33] The realization that such models of disorder used to represent spin disorder (in DLM) or chemical disorder in random alloys are lacking, motivated more recently[32-34] a more general approach allowing the charge and moments on each site to depend on its local environment. This is possible by using a supercell or by a multi-site description, as will be discussed below. This allows the magnetic moment on Eu to always be finite and to depend on its local coordination environment.

In light of the above discussion of the different views on the microscopic nature of the disorder, we wish to focus on two interesting features in the para phases of ETO:

*(a) Magnetic activity in the PM phases:* Although the PM phases of ETO were often referred to as "*nonmagnetic*" or "*nominally nonmagnetic*" [11,14], the PM phases are magnetically active in an external magnetic field as well as without field, developing small magnetic regions significantly above the Néel temperature.[9,12-14,35-38] This behavior was attributed to the spin-lattice coupling noted in the *spin-ordered AFM* phase based on the dependence of dielectric constant on the magnetic field[37] and DFT calculations of AFM-to-ferromagnetic transition on dielectric constant and phonon frequencies.[39] Whereas a similar scenario of spin-lattice coupling has also been suggested for the spin disordered PM phases[40] this is yet unclear given that the magnitude and distribution of the magnetic moments in the PM phases is unknown. As shown in Table I, provided one allows larger than conventional unit cells, the two PM phases are predicted by DFT to have a distribution of local magnetic moments of similar magnitude as the AFM phase.

*(b) Local structural symmetry breaking in the $\gamma$ PM phase:* A recent analysis of the diffraction measured pair distribution function (PDF)[2] demonstrated that the nominally cubic $\gamma$ phase manifests local octahedral tilting similar to those in low-temperature magnetically ordered tetragonal $\alpha$ phase. This result thus demonstrates that the crystallographic structure of the $\gamma$ PM phase is incompatible with the assigned nominal cubic Pm-3m symmetry having but a single octahedron per unit cell. In contrast, however, the PDF for the $\beta$ PM phase at 100 K can be explained with I4/mcm structure[2]. As Table I suggests, if the cubic $\gamma$ PM phase is described by a larger than a conventional phase, DFT calculation predicts not only a distribution of local magnetic moments but also a distribution of lattice octahedral tilting.

The foregoing discussion suggests that a paramagnetic phase could manifest a distribution of (mutually compensating) local moments. In addition, if the PM is made of octahedra (as in perovskites) that can tilt, the PM phase can also manifest a distribution of local lattice tilting and distortions. Thus, depending on the phase, the PM phase of perovskites can have both spin symmetry breaking as well as lattice symmetry breaking. This opens the door also for mutual coupling between spin and lattice. *Thus, paramagnetism and paraelasticity can coexist hand in hand as two forms of energy lowering local symmetry breaking.* Thus, we will abandon the tradition of assigning *minimal monomorphous unit cell* having a single untilted octahedron and a single spin motif to describe a PM phase. We also avoid the harmonic phonon as a starting point of view for coupling, using instead a full Born-Oppenheimer surface without specializing to small deviations from the well minimum. Instead, we allow a larger cell where both positional and spin symmetry breaking lower the energy. We find that the β PM phase retains the positional minimal tetragonal I4/mcm cell (no formation of the structural polymorphous network characterized by the existence of a distribution of positional local motifs) but shows magnetic symmetry breaking, characteristic of a larger polymorphous *spin* unit cell with nonzero Eu local magnetic moments. In contrast, the γ PM phase must be described using polymorphous (pseudo) cubic network simultaneously accounting for both positional and spin-broken symmetries. These results obtained by minimization of the quantum mechanical forces with allowing for local magnetic moments to develop finite values subject to the zero global moment condition (see DFT details) explain why β and γ PM phases are magnetically active and why they have an intrinsic tendency for structural/spin symmetry breaking.

## II. Results and discussion

### A. The α phase of EuTiO$_3$: antiferromagnetic tetragonal insulator

By screening the DFT calculated total energy of all unique magnetic configurations in tetragonal ETO supercells containing up to 8 formula units (fu), we identify that the α phase is an antiferromagnetic (AFM-G) insulator, in agreement with experimental observations[7] and other first-principles calculations[41-45].

*Local moment and local octahedral tilting in the α phase:* Figure 1a,b shows the crystal structure of the lowest energy AFM configuration of the α phase of ETO and the DFT calculated density of states. In the AFM magnetic configuration, each Eu atom has a DFT magnetic moment of 6.89μ, while magnetic moments on Ti and O are zero within numerical precision. In the resulted structure, Ti atoms do not have any off-centered displacements and TiO$_6$ octahedra exhibit a$^0$b$^0$c$^-$ tilting (i.e., R$_5^-$ distortion mode of Pm-3m structure) with an amplitude of 7.83° in agreement with that (~7.5°) predicted by other first-principles calculations[41]. This tilting angle is overestimated as compared to 3.53° corresponding to the experimentally observed crystal structure (experimental lattice vectors and Wyckoff positions) based on the Rietveld refinement against neutron powder diffraction data collected at 1.5 K[6]. The difference in rotation angle obtained with DFT is not due to difference in lattice constants: the experimental lattice constants are a = 5.50 and c = 7.80 Å[6], while the corresponding DFT (PBE+U) values are a = 5.55 and c=7.94 Å. When we use the experimental lattice constant rather than the DFT optimized value, the tilting angle is 8.1 degrees. We surmise that the smaller measured tilting angle is associated with the over localization of the Eu-4f electrons missed by DFT without good cancellation of the self-interaction energy. This picture is confirmed by fact that the increase of U values results in approaching the DFT TiO$_6$ tilting angle towards the experimental value.[41]

*Scenario of ferroelectricity in the α phase:* Recently, the observation of soft phonons in the α and β phase [46,47] was interpreted as ferroelectric-like behavior (i.e., Ti-off centering)[47]. We note that DFT total energy calculations were recently shown to reliably and systematically predict which compounds are ferroelectric compounds[48] and which are not. Using the same DFT we find no ferroelectric Ti displacements in equilibrium ETO in agreement with other first-principles simulations[39,41,42], and the crystallographic data identifying centrosymmetric space group for the α phase.[6] Recent experimental observation indicates that ETO is ferroelectric only under strain.[39,49]

*Electronic structure of the α phase:* The α phase of EuTiO$_3$ is an antiferromagnetic tetragonal insulator with PBE+U band gap energy of 0.33 eV (underestimated as compared to the experimental value of ~0.8-1 eV corresponding to the β and γ phases[15,50,51]). The upper valence band is composed of Eu-4f states with a minor contribution of Ti-3d and O-2p states and is occupied by 7e per formula unit. The deeper valence band is located at E$_v$-2 eV and is O-p like band. The conduction band is dominated by Ti-d states. Thus, the occupied narrow Eu-4f band is an isolated impurity-like band nested within the bonding-antibonding Ti-O gap. This computed electronic structure thus sheds light on the possible reason behind the 4.53 eV transition measured by spectroscopic ellipsometry at room temperature.[52] If the latter value corresponds to the minimum energy band gap then it contradicts existing theoretical and experimental literature on band gap energy of ETO phases. However, this value could correspond to the O-2p to Ti-3d transition as has been also suggested by the analysis of temperature-dependent optical absorption coefficient and its correlation to DFT electronic structure. Thus, the above transition measured by spectroscopic ellipsometry need not correspond to the minimum energy gap.

One may wonder what is the impact of local structure on the electronic properties of ETO. To answer this question, we compare the band gap energies for AFM-G α phase with forced 0 and equilibrium tilting. It turns out that octahedral tilting is not necessary for band gap opening in the α phase, moreover, the band gap energy for the untilted structure (i.e., 0.37 eV) is slightly larger than that (i.e., 0.33 eV) for the structure obtained by minimization of quantum forces. This small variation of band gap energy with respect to local internal structure is mainly caused by the uniqueness of ETO electronic structure where localization of Eu-4f states is weakly affected by TiO$_6$ octahedral tilting.

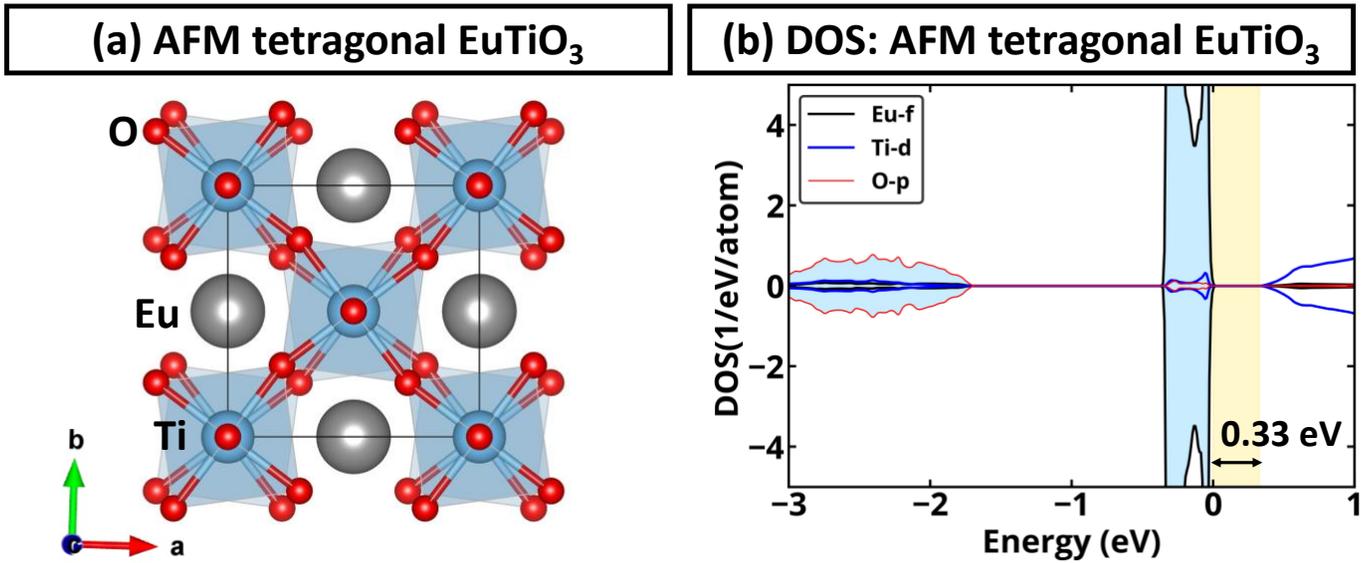

**Figure 1. The α phase of EuTiO$_3$: antiferromagnetic tetragonal insulator.** (a) Crystal structure of tetragonal EuTiO$_3$. (b) The orbital and atom projected density of states (DOS) for AFM-G type tetragonal EuTiO$_3$. The occupied states are shadowed in light blue. The upper valence band is Eu f-like, and the lower conduction bands are Ti-d like. The band gap region is shown in yellow. All results are presented for PBE+U calculations with a U-J value of 5.2 eV applied on Eu-f states.

### B. The β phase of EuTiO$_3$: paramagnetic tetragonal insulator

While the first-principles literature on ETO is rich of theoretical investigations of the magnetically ordered α[41-45] or cubic (not experimentally existing phase) Pm-3m[15,39,53] phases, the PM tetragonal ETO phase has not been discussed in the theoretical literature. To assess the polymorphous picture of ETO and establish if the symmetry breaking is only in the geometric atomic distortions ("positional") and/or magnetic ("spin"), we design calculations that can reveal this. The first allows for positional but not for spin symmetry breaking, whereas the second allows for both.

#### 1. The β PM phase of EuTiO$_3$ described with positional but not spin symmetry breaking

Figure 2a,b shows for the β phase (computed using a tetragonal (I4/mcm) primitive cell containing two ABO$_3$ formula units) the electronic structure, magnetic properties in the spin symmetry unbroken (i.e., nonmagnetic, non- spin-polarized) but positional symmetry broken description. While the main positional symmetry breaking is the R$_5^-$ distortion mode of Pm-3m as is the case for the α AFM phase, due to nonmagnetic assumption and un-paired *f* electrons in NM ETO, each atom has zero magnetic moment (Fig. 2b). This thus results in a (false) metal (Fig. 2a). Similar to Mott insulators, due to odd number of electrons, splitting of Eu states cannot be done by positional symmetry breaking alone. This thus suggests that ignoring magnetic symmetry breaking leads to the metallic state. This metallic electronic structure differs from that in typical monatomic metals (e.g., Cu or Al) in that the former has the Fermi level inside the principal conduction band but there is an internal wide band

gap below the conduction band minimum (CBM) and above the deep valence band maximum (VBM) (see Fig 2a). Compounds with such electronic structures can be thought of as "degenerate gapped metals".[54,55] In practice, they are often *False Metals* when the freedom to lower their energy by moving the Fermi level from the continuum into the principle VBM-CBM band gap region is not afforded in the calculation, as discussed in Ref.[23] Herein, we find that the symmetry unbroken nonmagnetic model results in DFT total energy that is 1.92 eV/atom above that of the α phase. This energy difference is enormously large as compared to the typical energy difference between low- and high-temperature phases of different compounds (e.g., in order of a few meV/atom for typical nonmagnetic $ABO_3$ perovskites[56,57]) clearly implying that the non-magnetic model is not a reasonable starting model.

### 2. The $\beta$ PM phase of EuTiO$_3$ described with both positional and spin symmetry breaking

To describe the properties of the PM tetragonal phase, we adopt the recently proposed spin polymorphous model[22-27,58] allowing for positional and spin symmetry breaking. All calculations are performed using the 2×2×2 supercell (32 fu/cell) of the nominal tetragonal I4/mcm structure. To allow for a local spin moment, we initially occupy Eu sites in the supercells by spin up and spin down so as to create a global zero moment configuration closest to the high-temperature limit of a random spin paramagnet. To establish the best simulation of random spin-statistics possible within a given supercell size, we use the special quasi-random structure (SQS)[34] to identify the occupation pattern with the smallest deviation between actual spin-spin correlation functions and those corresponding to *ideal random statistics (i.e., high-temperature limit of paramagnet)*. Simulation of a supercell with finite short-range order (SRO) is also possible by selecting site occupation numbers best matching a given spin-spin SRO. To enable local positional symmetry breaking such as octahedral tilting, should it lower the energy, we minimize quantum mechanical forces while restricting the supercell shape to the global macroscopic tetragonal lattice (for the γ phase discussed below, the supercell shape is restricted to the global macroscopic cubic lattice) after introducing random atomic displacements for each atom with arbitrary direction and maximum displacement amplitude of 0.1 Å.

Figure 2c,d shows the results for the β phase allowing for both spin and structural symmetry breaking (i.e., computed using the spin polymorphous model), demonstrating that the resulting system is a magnetic insulator with PBE+U band gap energy of 0.31 eV and internal DFT energy of over 1.91 eV/atom lower than that of the nonmagnetic symmetry unbroken model. Here, while the total magnetic moment is 0, there is a narrow distribution of local magnetic moments, i.e., each Eu site has an absolute local magnetic moment of 6.88±0.005μ. These results thus highlight the fundamental difference in the description of PM compounds with spin polymorphous model and nonmagnetic approach, demonstrating that accounting for a spin and structural

symmetry breaking can allow describing gapping in the β phase. Moreover, since each site in such PM tetragonal ETO structure has a local magnetic moment, it is not surprising that the β phase is magnetically active and responds to the magnetic field. While breaking the spin symmetry results in coupling to local structural symmetry breaking, for the β phase, such coupling is extremely weak and maximum Eu and Ti atomic displacements are less than 0.001 Å (i.e., numerically zero) with respect to their ideal Wyckoff positions. Breaking of local spin symmetry does not result in a significant change of local octahedral tilting as well – the amplitude of a$^0$b$^0$c$^-$ tilting is 8.01±0.02°, which is close to the value given above for the α phase (this angle is overestimated as compared to experimental studies as noted above). We still identify the R$_5^-$ distortion mode as the main symmetry breaking present in the β phase in spin polymorphous calculations. To estimate the relative contribution ($I_k$) of the R$_5^-$ distortion mode to symmetry breaking in the β phase, we calculate $I_k = \frac{A_k}{\sum_i^N A_i}$, where $A_i$ is the amplitude of *i* symmetry breaking mode observed in the system among N observed modes. The computed results suggest that the relative contribution of R$_5^-$ distortion mode is 99%, which is consistent with the fact that experimentally the β phase is known to have the I4/mcm symmetry without distinct structural symmetry breaking as confirmed by experimental measurements of the pair distribution function.[2]

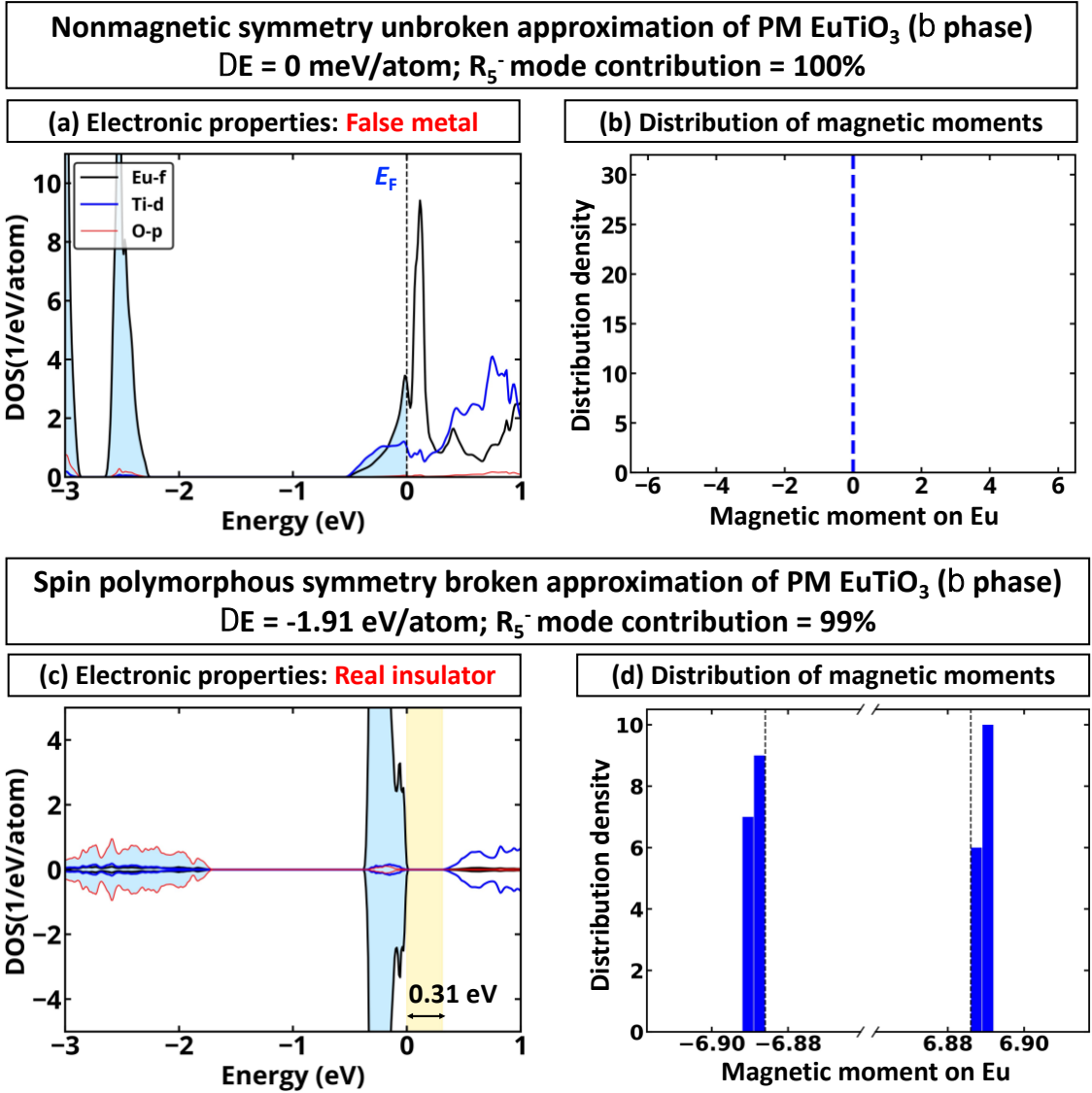

**Figure 2. The β phase of EuTiO₃: paramagnetic tetragonal insulator.** (a,c) Electronic and (b,d) magnetic properties of the paramagnetic β phase of EuTiO₃ computed for (a,b) nonmagnetic symmetry-unbroken model and (c,d) spin polymorphous symmetry-broken model. Spin polymorphous symmetry-broken model is reproduced by the nudging of 160-atom SQS supercell of tetragonal EuTiO₃ allowing internal relaxation and volume optimization but keeping tetragonal lattice vectors. The band gap region in (c) is shown in yellow. The dashed black line in (d) shows the magnetic moments in the α AFM phase. The contribution of different modes to symmetry breaking in the paramagnetic β phase is computed as $I_k = \frac{A_k}{\Sigma_i^N A_i}$, where $A_i$ is the amplitude of $i$ symmetry breaking mode observed in the system, and N is the number of symmetry-breaking modes present in the system with respect to Pm-3m structure. All results are presented for PBE+U calculations with a U-J value of 5.2 eV applied on Eu-f states.

### C. The γ PM phase of EuTiO₃: paramagnetic cubic insulator

#### 1. The γ PM phase of EuTiO₃ described with positional and spin unbroken symmetry

Figure 3a-c shows the results for electronic, magnetic, and symmetry for a nonmagnetic description of the γ PM phase. Similar to the case of the β phase, nonmagnetic (symmetry unbroken) approximation of the γ phase results in a degenerate gapped metal with Fermi level in the conduction band (Fig. 3a), which is in contradiction

with the experimental observation of an insulating state for the γ phase. The energy of nonmagnetic cubic ETO is 1.81 eV/atom above that for the ground state structure (α phase), i.e., it lies extremely in high energy and therefore unlikely to be of physical importance. Moreover, similar to the β phase, zero magnetic moment on each atom (Fig. 3b) of the monomorphous nonmagnetic system cannot explain why the γ PM phase is magnetically active above the Néel temperature.[9,12-14,35-37] Finally, such a model (Fig. 3c) does not allow to explain why the experimental pair distribution function revealed that the γ phase exists as the symmetry-broken structure with local structural motifs of the α phase.[2]

### 2. The γ PM phase of EuTiO$_3$ described with both positional and spin symmetry breaking

Figure 3d-f shows the results of the application of the spin polymorphous model to 2√2×2√2×4 supercell (32fu/cell) of nominal Pm-3m ETO structure. In contrast to the nonmagnetic monomorphous approximation of the γ PM phase, allowing spin and structural symmetry breaking results in a structural and spin polymorphous system with the PBE+U band gap energy of 0.27 eV (Fig. 3d) and a distribution of local magnetic and structural motifs (Fig. 3e,f). First, while the total magnetic moment of the cell is 0, the Eu sublattice has a small distribution of local magnetic motifs on Eu atoms with an average absolute value of the magnetic moment of 6.88±0.005μ. These results thus suggest that spin polymorphism does not result in substantial distribution of local magnetic moments on magnetic species, however, this is not always the case for PM compounds. For instance, PM monoclinic YNiO$_3$[59] and PM tetragonal FeSe[60] have a significantly larger variation of magnetic moments on metal sublattices. The obtained local moments on each Eu ion thus clearly implies that the γ phase is magnetically active, which is in agreement with experimental observations.[9,12-14,35-37] Second, the resulting structure has different structural motifs (see Fig. S1): there are (i) distribution of small atomic displacements for Eu and Ti atoms and (ii) distribution of octahedral tilting as compared to the ideal monomorphous Pm-3m ETO structure. While the maximum Eu and Ti displacements in the structure are 0.08 and 0.03 A, respectively, the maximum averaged displacement (i.e., $<x>=\frac{1}{N}\sum_{i=0}^{N}x_i$, where N is a number of corresponding atoms) along a, b, and c axis for Eu and Ti atoms is less than 0.001 Å for a 160-atom cell. To further understand the relation of the symmetry breaking observed in the spin polymorphous model, we apply the analysis of local structural symmetry-breaking modes as compared to the monomorphous Pm-3m structure. The results are summarized in Fig. 3f, showing that: (i) the γ phase exhibits the distribution of different structural symmetry-breaking modes as compared to Pm-3m structure; (ii) the main dominant mode corresponds to R$_5^-$, which is the same as that present in the AFM tetragonal α phase, (iii) T$_2$, DT$_5$, and M$_5^+$ are other symmetry-breaking modes with noticeable amplitude observed in the polymorphous structure. These data thus confirm

that PM cubic ETO structure tends to minimize the energy by structural symmetry breaking adapting the motifs (i.e., $R_5^-$ distortion) of the low-temperature α phase, which is in good agreement with results predicted based on the analysis of the diffraction measured pair distribution function[2]. These results thus also conclude that *the γ phase cannot be described as ideal high-symmetry cubic perovskite structure and indeed exhibit both structural and spin m-DOFs*, which only can be captured using a non-trivial supercell allowing for spin and structural symmetry breaking.

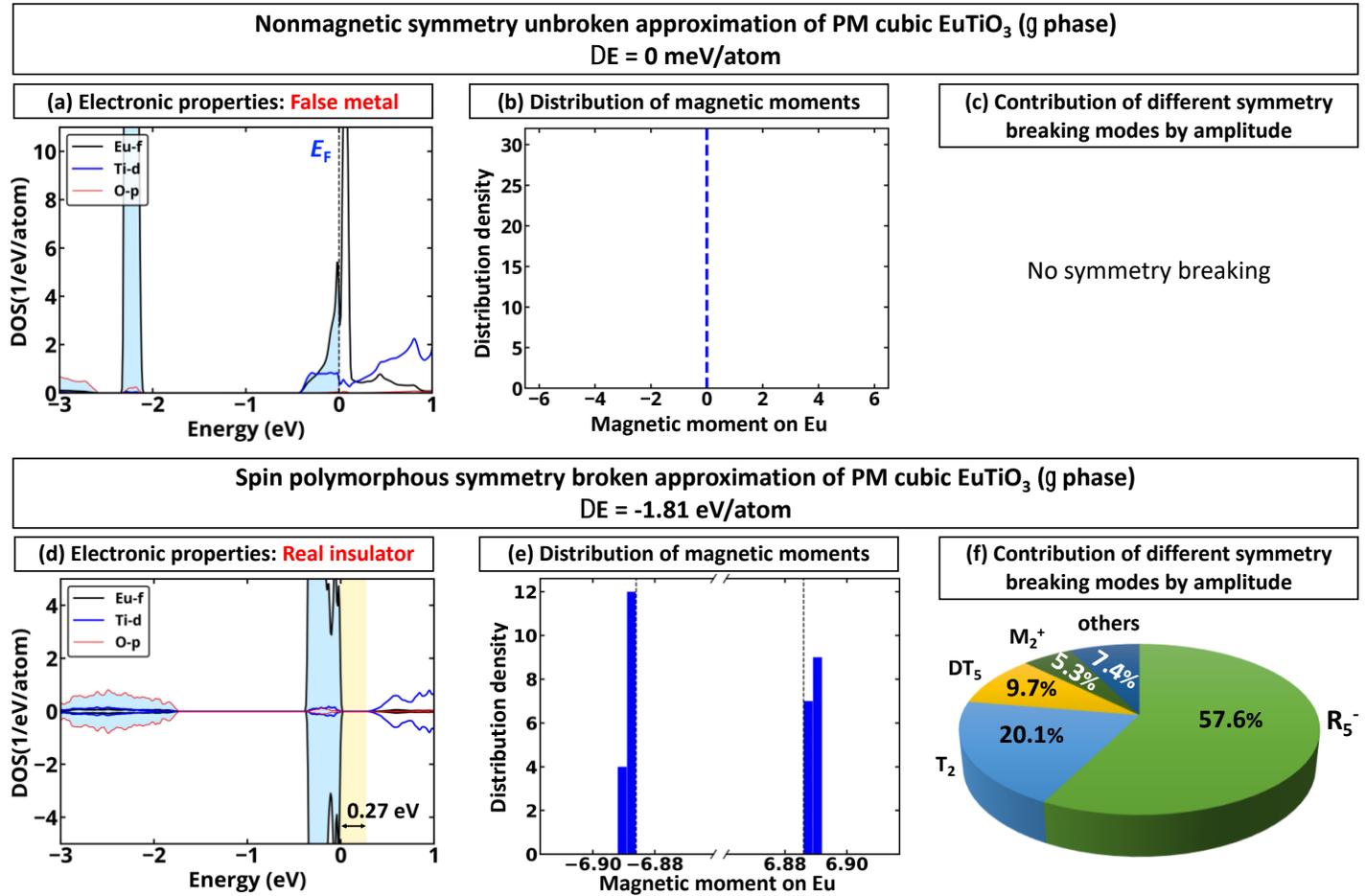

**Figure 3. The γ phase of EuTiO$_3$: paramagnetic cubic insulator.** (a,d) Electronic, (b,e) magnetic, and (c,f) structural properties of paramagnetic γ phase of EuTiO$_3$ computed for (a-c) nonmagnetic symmetry-unbroken model and (d-f) spin polymorphous symmetry-broken model. The spin polymorphous symmetry-broken structure is obtained by the nudging of 160-atom SQS supercell of cubic EuTiO$_3$ allowing internal relaxation and volume optimization but keeping cubic lattice vectors. The band gap region in (d) is shown in yellow. The dashed black line in (e) shows the magnetic moments in the α phase. Contribution of different modes to symmetry breaking in paramagnetic γ phase is computed as $I_k = \frac{A_k}{\Sigma_i^N A_i}$, where $A_i$ is the amplitude of *i* symmetry breaking mode observed in the system, and N is the number of symmetry-breaking modes present in the system with respect to Pm-3m structure. All results are presented for PBE+U calculations with U-J value of 5.2 eV applied on Eu-f states.

## III. Conclusions

Allowing for the existence of different *structural local motifs* (rotated, tilted, deformed or disproportionated $BO_6$ octahedra) and *spin local motifs*, we demonstrate that α, β, and γ phases of ETO develop different degrees of symmetry breaking. At low temperatures, the α phase - magnetically ordered AFM tetragonal ETO is an insulator that exhibits $R_5^-$ ordered distortion with respect to the Pm-3m structure and each Eu atom having the same magnetic moments. This α phase can be described using a small primitive cell containing only 4 fu accounting for a single distortion mode. The β phase is a PM tetragonal insulator that is spin polymorphous and exhibits magnetic symmetry breaking with each Eu atom being magnetically unique and having its local magnetic moment. The accurate description of this phase (as well as γ phase) requires using large supercell with allowing spin symmetry breaking (e.g., via employing special quasi-random structure spin distribution). In the resulting symmetry broken structure, the β phase still exhibits the $R_5^-$ distortion but does not have other noticeable structural symmetry breaking modes. The high-temperature PM cubic $EuTiO_3$ (γ phase) is a polymorphous insulator that exhibits both structural and magnetic symmetry breaking as a result of internal energy minimization. The γ phase has the distribution of both local spin and structural motifs. Here, the $R_5^-$ distortion remains the main structural symmetry-breaking mode, suggesting that the internal structure of the γ phase mimics the distortion observed in the α phase, with some contribution of other symmetry breaking modes that are not present in the low-temperature α phase, which is in good agreement with experimental results. Importantly, we demonstrate that in properly described PM β and γ phases, each Eu atom has local magnetic moments, which thus allows to explain why experimentally one observes local magnetic activity in $EuTiO_3$ significantly above the Néel temperature and why there is a magnetic field dependence of β-γ transition temperature.

**Acknowledgment:** U.S. Department of Energy, Office of Science, Basic Energy Sciences, Materials Sciences and Engineering Division within DE-SC0010467 support this work. The authors acknowledge the use of comp Extreme Science and Engineering Discovery Environment (XSEDE) supercomputer resources, which are supported by the National Science Foundation, grant number ACI-1548562.

## APPENDIX: DFT details

The first-principles calculations are performed using pseudopotential plane-wave DFT as implemented in the Vienna Ab Initio Simulation Package (VASP)[61-63] with Perdew-Burke-Ernzerhof (PBE)[64] exchange-correlation functional and +U correction (U-J value of 5.2 eV) introduced by Dudarev et al.[65] applied on Eu-f states. The cutoff energies for the plane-wave basis are set to 500 eV for final calculations and 550 eV for volume relaxation. Atomic relaxations are performed until the internal forces are smaller than 0.01 eV/Å, unless specified. To identify the main the symmetry-breaking modes in each phase, we employed the AMPLIMODES[66,67] code that allows identifying the symmetry-breaking modes in the compound via generating atomic displacement patterns induced by irreducible representations of the parent (i.e., Pm-3m in this work) space-group symmetry.

APPENDIX: Distribution of local structural motifs in the $\gamma$ phase of EuTiO$_3$ as computed using the spin polymorphous model

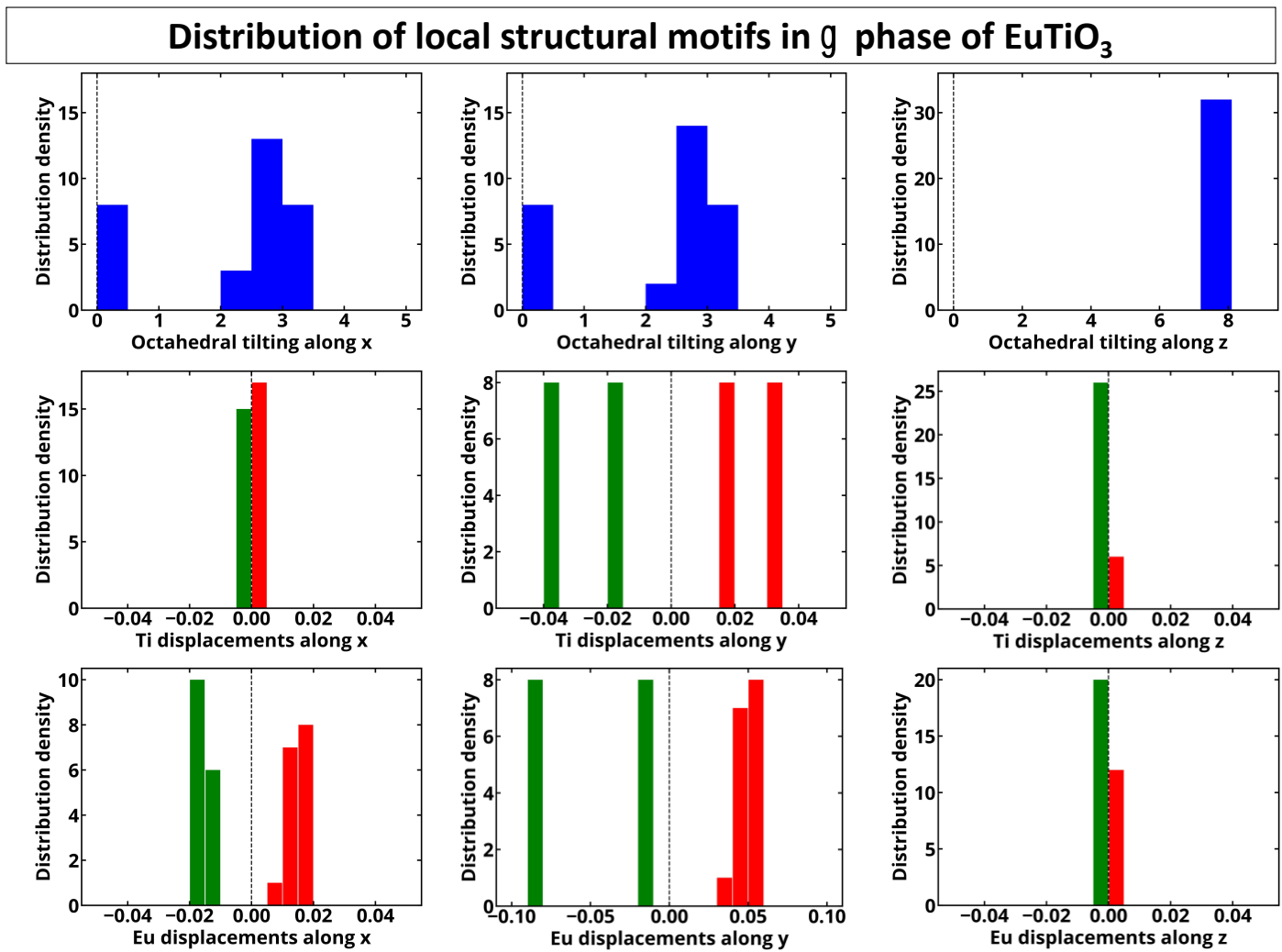

**Figure S1.** Distribution of local structural motifs in spin and structural polymorphous cubic EuTiO$_3$ shown as distribution of octahedral tilting, Eu displacements, and Ti displacements. The maximum averaged displacement (e.g., $<x> = \frac{1}{N}\sum_{i=0}^{N} x_i$, where N is number of corresponding atoms) along x, y, and z axis for Eu and Ti atoms

is less than 0.001 Å for 160-atom cell. Corresponding tilting angles and Eu/Ti atomic displacements in monomorphous cell are shown by dashed line.